\documentclass[twocolumn,noshowpacs,preprintnumbers,amsmath,amssymb,aps]{revtex4-2}
\usepackage{graphicx}% Include figure files
\usepackage{dcolumn}% Align table columns on decimal point
\usepackage{bm}% bold math
\usepackage[usenames]{color}
\usepackage{colortbl}
\usepackage[utf8]{inputenc}

\begin{document}

\title{Word and bit line operation of a $1\times 1~\mu m^2$ superconducting vortex-based memory}
%\title{Demonstration of a multibit vortex-based superconducting memory}
\author{Taras Golod}
\author{Lise Morlet-Decarnin}
\author{Vladimir M. Krasnov}
\email{vladimir.krasnov@fysik.su.se}

\affiliation{Department of Physics, Stockholm University,
AlbaNova University Center, SE-10691 Stockholm, Sweden.}
%\date{November 2022}

\begin{abstract}
%A paradigm shift from semiconductors to superconductors could revolutionize digital computation techniques. 
The lack of dense random access memory is one of the main bottlenecks for the creation of a digital superconducting computer. In this work we study experimentally vortex-based superconducting memory cells. Three main results are obtained. First, we test scalability and demonstrate that the cells can be straightforwardly miniaturized to submicron sizes. Second, we emphasize the importance of conscious geometrical engineering. In the studied devices we introduce an asymmetric easy track for vortex motion and show that it enables a controllable manipulation of vortex states. Finally, we perform a detailed analysis of word and bit line operation of a $1\times 1 ~\mu$m$^2$ cell. High-endurance, non-volatile operation at zero magnetic field is reported. Remarkably, we observe that the combined word and bit line threshold current is significantly reduced compared to the bare word-line operation. This could greatly improve the selectivity of individual cell addressing in a multibit RAM. The achieved one square micron area is an important milestone and a significant step forward towards creation of a dense cryogenic memory.   
\end{abstract}
\maketitle

\section{Introduction}

%Transition to sustainable development is a global challenge, facing our society. Effective
%utilization of electrical energy is a key factor for decarbonization of economy. 
The development of superconducting (SC) electronics could lead to a breakthrough in future computation techniques. Major advances in the creation of a SC quantum computer were recently achieved \cite{Arute_2019,Kandala_2019}. However, practical calculations today are made on a classical computer and demands for
digital computation capacities are growing in the explosive manner. Resistivity causes principle limitations for semiconductor-based electronics.
The large resistance, $R$, of silicon transistors both limits the operation speed (determined by the $RC$ time constant) 
%, where $C$-is the capacitance), 
and creates a problem of heat management in very-large-scale-integration (VLSI) circuits. Those obstacles can be obviated by shifting to superconductors with $R = 0$. The maximum operation frequency of SC electronics is determined by the energy gap, which can exceed 10 THz in high-temperature SCs \cite{Borodianskyi_2017}. 
%This has to be compared to a few GHz frequency in semiconducting computers. 
For large data facilities, shifting from semiconductors to superconductors could drastically improve both the power efficiency (by an order of magnitude) and the computation speed
(by several orders of magnitude). Such perspectives reignited the interest to a classical SC computer \cite{Holmes_2013,Ortlepp_2014,Manheimer_2015,Golod_2015,Soloviev_2017,Herr_2018,Tolpygo_2019,Semenov_2019,Chen_2020}.

Digital SC electronics has a long history. Rapid-single-flux-quantum (RSFQ) architecture was developed almost half a century ago \cite{Likharev_1986}. It is based on storage and manipulation of the flux quantum, $\Phi_0$, in superconducting
quantum interference devices (SQUIDs). %, usually containing two JJs. 
Nb-based RSFQ electronics is capable of operation up to $\sim 200$ GHz frequency \cite{Holmes_2013}, two orders of magnitude faster than modern computers. However, RSFQ
has a major problem with scalability. The current needed for introduction of $\Phi_0$ is
determined by the inductance, $L$, of the SQUID loop, $I = \Phi_0/L$. Upon miniaturization, $L$ decreases and the operation current increases inversely proportional to the size. 
Therefore, RSFQ is not compatible with the VLSI technology. The main bottleneck is the lack of dense random-access memory (RAM) \cite{Holmes_2013,Ortlepp_2014,Manheimer_2015}. State of the art RSFQ RAM has a footprint of $\sim 100~\mu m^2$ per bit \cite{Nagasawa_2006,Ortlepp_2014,Semenov_2019,Chen_2020}. 
Large sizes cause significant delay times. %for propagation of control pulses. 
In fact, the speed of RSFQ %electronics 
is limited by delay times, rather than the energy gap  \cite{Holmes_2013}.
%The growing demand for a power-efficient exaflop supercomputer has led to renewed efforts for development of a digital SC computer. In 2013 a large project has been initiated in USA under the Intelligence Advanced Research Projects Activity program [14]. The key
%focus of this project is on development of scalable cryogenic RAM. Although the final output
%of this project is not yet publicly known, a proposed 

Novel approaches are needed for building a VLSI-competitive SC electronics. Several strategies for making dense SC RAM have been suggested recently  \cite{Hilgenkamp_2006,Zdravkov_2013,Vernik_2013,Goldobin_2013,Baek_2014,Golod_2015,Bezryadin_2017,Birge_2018,Nevirkovets_2018,Birge_2019,Prada_2019,Giazotto_2021,Karelina_2021,Berggren_2021,Alam_2021,Aarts_2022}. %, many of which are awaiting for experimental approval. 
In Ref. \cite{Golod_2015} it was shown that a single Abrikosov vortex (AV) can be used as an information carrier. AV represents the smallest magnetic object in a superconductor, %, with the size determined by the London penetration depth, $\lambda_L\sim 100$ nm. 
enabling miniaturization to submicron sizes. %The AV can be easily manipulated and detected \cite{Golod_2010,Golod_2019,Golod_2021}. 
%The quantized nature of AV facilitates persistent, non-volatile and high-endurance operation of AVRAM (AV-based RAM). 
Prototypes of single-bit AVRAM cells with excellent performance %half-selection stability of at zero magnetic field has 
were demonstrated \cite{Golod_2015}. %. The anticipated time, $\tau \lesssim 1$ ns, and energy per operation, $E \lesssim 10^{-18}$ J, satisfying stringent demands for a cryogenic RAM \cite{Ortlepp_2014,Manheimer_2015}. 
However, the way from a single cell to a dense multi-bit RAM is full of hazards, as could be learned from the history of MRAM \cite{Parkin_2006}. Cross-talking between cells could hamper RAM operation. It could be mitigated by a highly selective word and bit line (WL \& BL) addressing of individual cells. %in a multi-bit RAM. 
So far WL \& BL operation of AVRAM cells has not been analyzed. 

In this work we study experimentally Nb-based AVRAM cells. Our aim is threefold. First, we study scalability. Cells with the same geometry, but different sizes 5, 3 and $1~\mu$m are tested. We confirm that AVRAM cells are scalable to submicron sizes, while keeping robust non-volatile operation at zero magnetic field. Second, we emphasize the importance of conscious geometrical design for the optimization of cells. In the studied cells, we introduce an asymmetric ``easy track" for vortex motion. % at one side of the vortex trap.  We show that 
It removes the degeneracy between vortex and antivortex states and, thus, facilitates controllable operation of the cell. Finally, we analyze the word and bit line operation of a $ 1\times 1~\mu$m$^2$ cell. We observe that simultaneous application of WL and BL pulses can significantly reduce threshold currents for vortex manipulation. %, compared to the bare WL operation. %Threshold currents as small as few tens of $\mu$A are achieved. 
Such a cooperative WL \& BL effect could greatly improve the selectivity of addressing in a multibit RAM. We conclude that AVRAM is a feasible candidate for the creation of dense cryogenic memory.

\begin{figure*}[]
    \centering
    \includegraphics[width=0.99\textwidth]{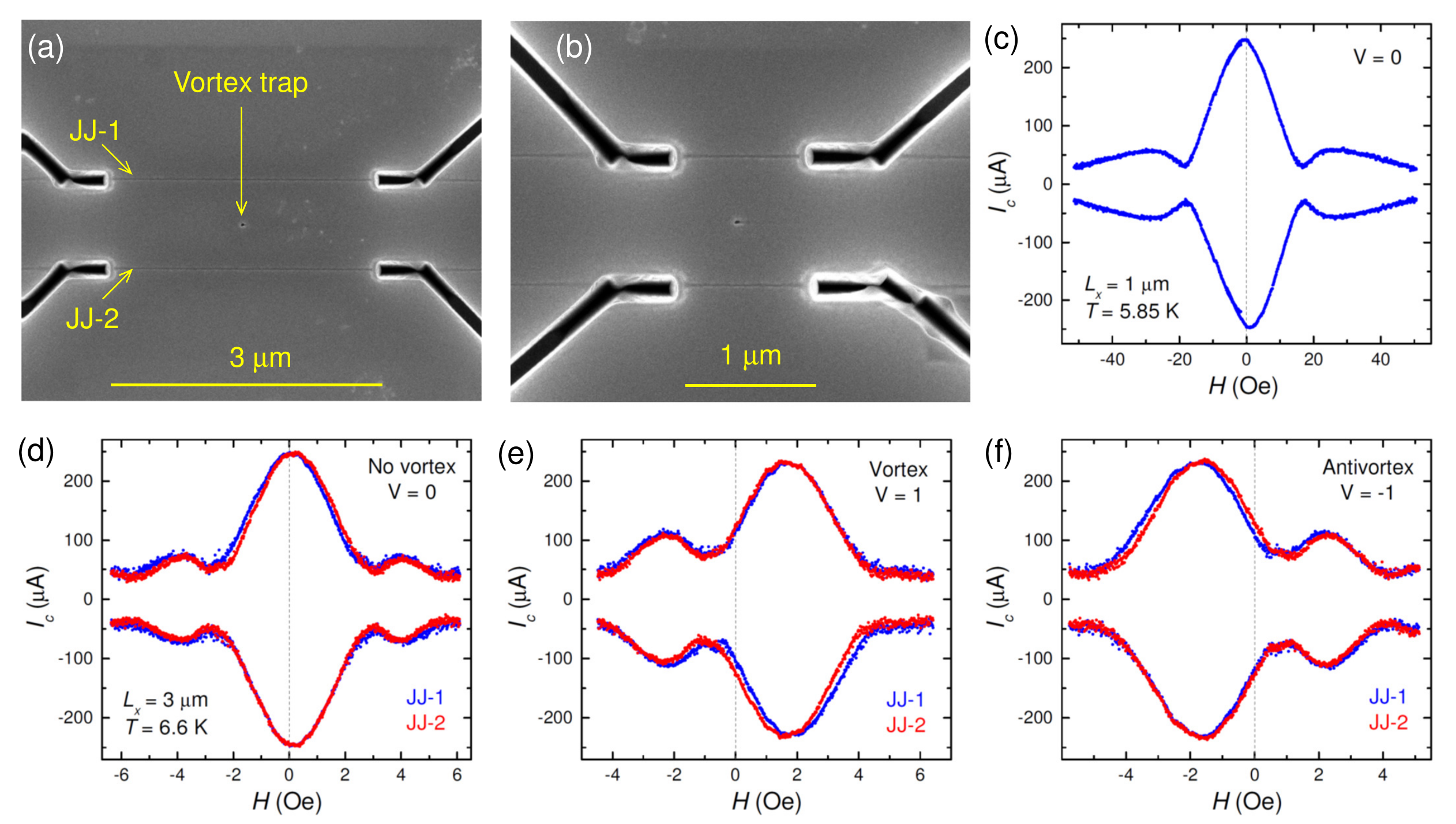}
    \caption{ {\bf Demonstration of AVRAM cell scalability.} (a) and (b) SEM images of two cells with sizes (a) $L_x \simeq 3~\mu$m, and (b) $L_x\simeq 1~\mu$m. (c) Magnetic field modulation of the critical current in a readout junction on the $L_x=1~\mu$m cell in the vortex free state, V=0. (d-f) $I_c(H)$ modulations for both junctions on the $L_x=3~\mu$m cell: (d) in the vortex-free state, V=0, (e) with a trapped vortex, V=1, and (f) with an antivortex V=-1.    
    }
    \label{fig:fig1}
\end{figure*}

\section{Results}

\subsection{Samples}

Planar AVRAM cells %, similar to Ref. \cite{Golod_2015}, 
are made from a single Nb film of thickness $d= 70$ nm. The film was deposited on an oxidized Si substrate by magnetron sputtering. First, a $\sim 5~\mu$m cross-like structure was made by photolithography and reactive ion etching. Subsequently, the sample was transferred into a dual-beam $^+$Ga Focused Ion Beam (FIB) and 
%system (FEI-Nova200). 
two planar readout Josephson junctions (JJs), of variable-thickness type, %, seen as parallel horizontal lines in the central parts of the images, 
were made by cutting narrow grooves in Nb. 
%The JJs are of the variable-thickness (Nb-constriction-Nb) type. The cuts were made as single-pixel lines with the nominal etching depth of 60 nm for the standard Si-milling parameters at an ion current of 10 pA (30 keV). 
Finally, a vortex trap, a nano-scale hole, was made in the electrode between the JJs. More details about junction fabrication and characterization can be found in Refs. \cite{Golod_2015,Golod_2021,Golod_2022,Grebenchuk_2022}. The fabrication procedure is highly reproducible and the characteristics of both JJs are practically identical. Several devices with similar geometry but different lengths of readout JJs, $L_x = 5,~3$ and $1~\mu$m, were made simultaneously on the same chip. Measurements are performed in a closed-cycle cryostat. Magnetic field, perpendicular to the Nb film was supplied by a superconducting magnet. %The $T_c$ of Nb film is $\simeq 8.8$??? K. 
We can controllably introduce and remove vortices by short current pulses, as described in Refs. \cite{Golod_2015,Supplem}. 
Additional information can be found in the Supplementary material \cite{Supplem}. 

\subsection{Miniaturization}
%To study the scalability, 
Figures \ref{fig:fig1} (a) and (b) show scanning electron microscope (SEM) images of the cells with (a) $L_x = 3~\mu$m and (b) $L_x = 1~\mu$m. Figs. \ref{fig:fig1} (c) and (d) show magnetic field dependencies of critical currents, $I_c(H)$, for the two devices in the vortex-free state, V=0.  All JJs exhibit Fraunhofer-type modulation with the central maximum at $H=0$. The flux quantization field for the $1~\mu$m device (c) is approximately 7 times larger than for the $3~\mu$m device (d), consistent with the expected difference in flux quantization areas \cite{Grebenchuk_2022,Supplem}. 
The largest cells, $L_x = 5~\mu$m, are similar to those, studied in Ref. \cite{Golod_2015}. Their characteristics can be found in the Supplementary \cite{Supplem}. 

\begin{figure*}[]
    \centering
    \includegraphics[width=0.95\textwidth]{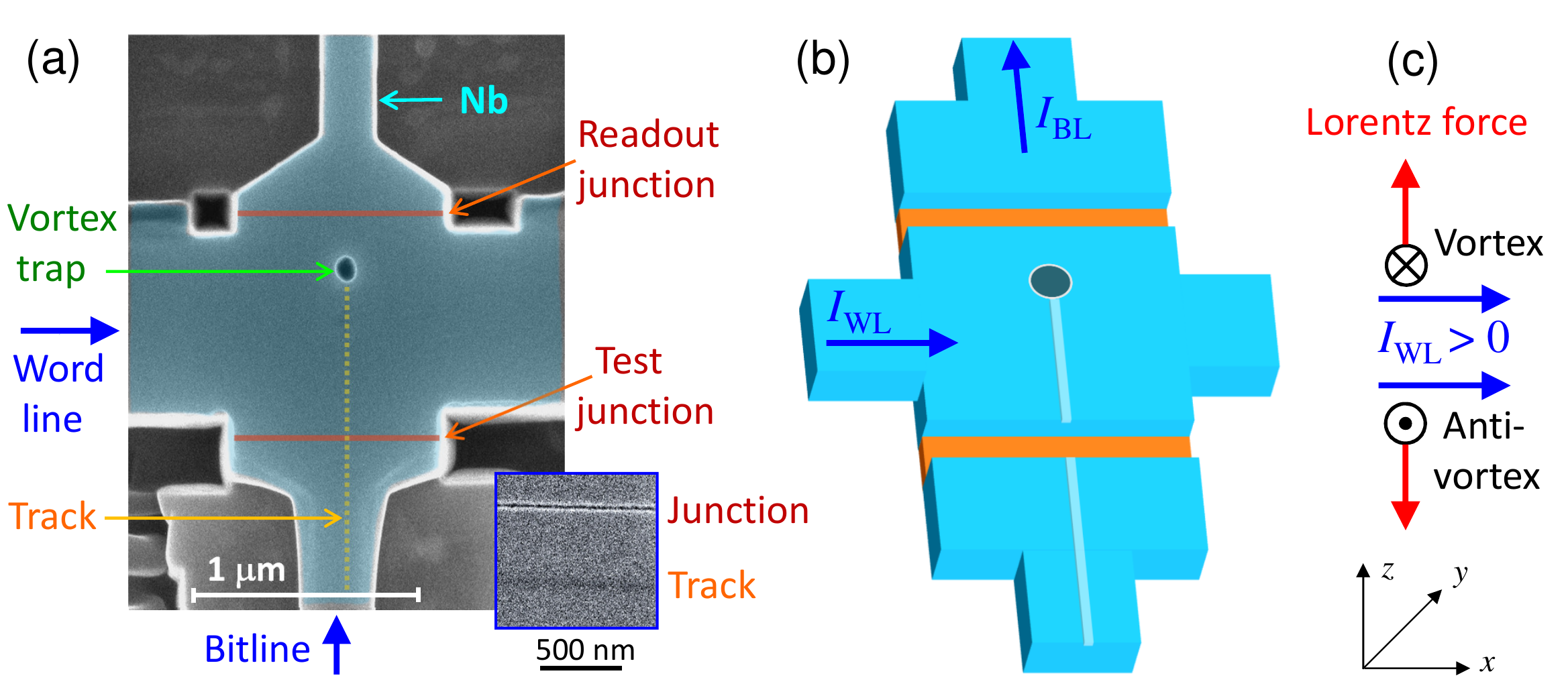}
    \caption{ {\bf Geometry of a $1\times 1 ~\mu$m$^2$ cell with word and bit lines.} (a) SEM image of the device (false color). Inset shows a SEM image of a junction groove and a FIB track for another device. (b) A sketch of the cell with indicated directions of wordline and bitline currents. (c) Direction of Lorentz forces, exerted on a vortex and an antivortex by a positive WL current. The easy operation with a small WL threshold current is expected when the vortex is transported along the track. 
    }
    \label{fig:fig2}
\end{figure*}

In Figs. \ref{fig:fig1} (e) and (f) we show $I_c(H)$ patterns for both readout JJs on the 3 $\mu$m cell with a trapped vortex (e), or an antivortex (f). From Figs. \ref{fig:fig1} (d-f) it is seen that the three primary states with different vorticities, (d) V=0, (e) V=1 and (f) V=-1, are clearly distinguishable. Vortex trapping leads to a characteristic distortion of $I_c(H)$ 
%The central maximum is displaced to a negative field for the vortex (e), and to the positive field for the antivortex (f) 
\cite{Golod_2010,Golod_2015,Golod_2019,Krasnov_2020,Golod_2021}, and to a threefold reduction of %the critical current at zero field. 
$I_c(H=0)$. 
This enables a simple recognition and readout of vortex states. The closer the AV is to the JJ, the larger is the junction response \cite{Golod_2019}. In these cells, the vortex trap is placed symmetrically with respect to the JJs; see Fig. \ref{fig:fig1} (a). Consequently, the responses of both JJs are identical and $I_c(H)$ patterns for both JJs merge in Figs. \ref{fig:fig1} (e) and (f). Such a coincidence confirms that the vortex is indeed placed in the trap. We always use two JJs in our AVRAM cells: one for determining the state (the readout JJ) and the other for confirming that the vortex is located in the trap (the test JJ). 

In Figures \ref{fig:fig2} (a) and (b) we show a SEM image and a sketch of a $\sim 1\times1~\mu$m$^2$ cell, in which excessive parts of the Nb electrode were removed. The trap with a diameter of $\sim 100$ nm is placed close ($\sim 240$ nm) to the top readout JJ. The bottom, test JJ is at a significantly longer distance ($\sim 740$ nm). %The response signal of this JJ is much smaller because of the larger distance to the trap. Therefore, 
%We will not show data for the test JJ, but just ensure that all states described below correspond to a vortex placed in the trap. 
The left-right and the bottom-top electrode pares form WL and BL, respectively. Below we will focus on the analysis of this cell. %Additional data for other devices can be found in the Supplementary \cite{Supplem}.

\subsection{Geometrical asymmetry}  
%Easy track for vortices and WL/BL asymmetry}

Vortex states can be manipulated by current pulses \cite{Golod_2015}. However, in a symmetric cell, 1 and -1 states are degenerate at $H=0$. Currents of any direction create both a vortex and an antivortex at opposite sides of the device and drive them inside, where they annihilate. Therefore, operation is impossible in a perfectly symmetric cell. In reality, there is always some asymmetry that lifts the degeneracy. However, the operation can not be fully controllable without conscious geometrical design. 

The asymmetry in the device from Fig. \ref{fig:fig2} (a) is introduced by making an easy track for vortex motion at one side of the trap. It is made by a single %back-and-forth 
scanning of the FIB to/from the trap. The inset in Fig. \ref{fig:fig2} (a) shows a high-resolution SEM image of the track alongside of the junction groove (for another device). It is seen that the track makes just a minor depression in the Nb film. It does not form a JJ, but %slightly suppresses superconductivity and 
creates an easy path for AV to and from the trap \cite{Golod_2021}. Fig. \ref{fig:fig2} (c) indicates directions of Lorentz forces, $F_L$, exerted on a vortex and an antivortex by a positive WL current. %, $I_{\text{WL}}>0$. %$F_L=(d/c) J\times \Phi_0$, where $J$ is the current density. 
From the comparison with the track position in Fig. \ref{fig:fig2} (a), it is expected that a positive $I_{\text{WL}}$ would easily introduce a vortex ($0 \rightarrow 1$) and remove an antivortex ($-1 \rightarrow 0$) along the easy track. All other operations should be more difficult and require larger threshold currents. 

\begin{figure*}[t]
    \centering
    \includegraphics[width=0.9\textwidth]{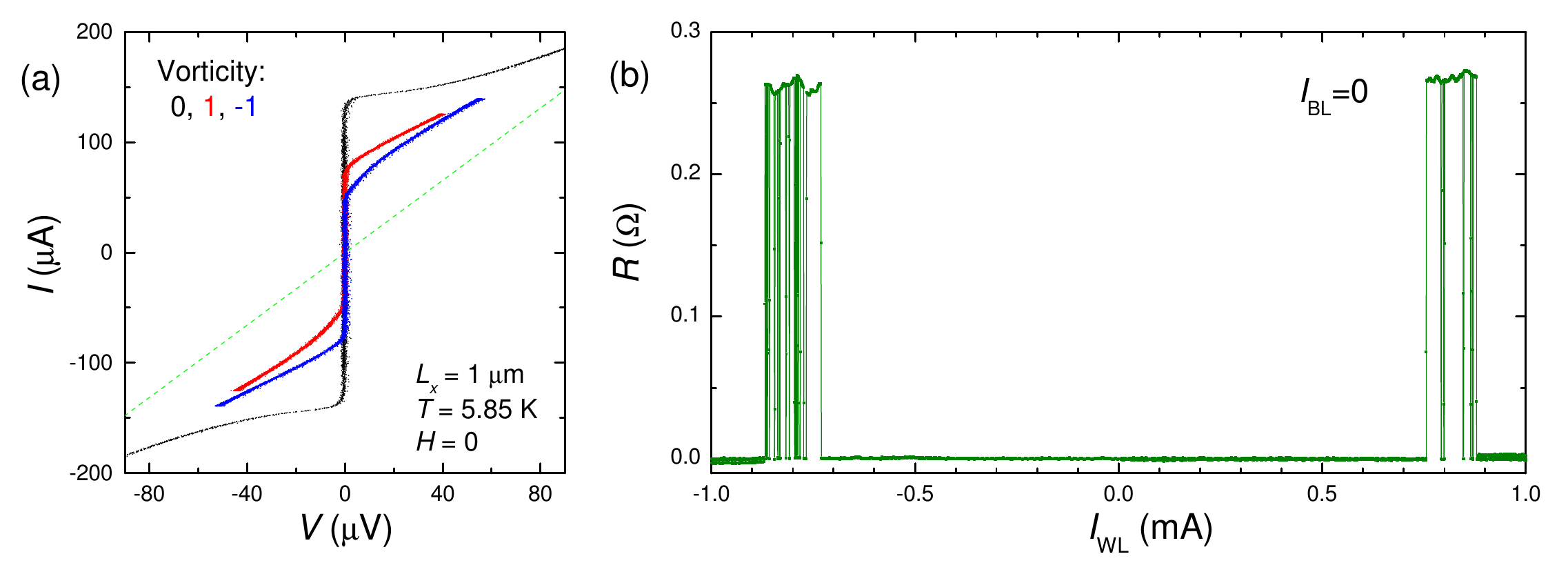}
    \caption{ {\bf Characteristics of the $1\times 1~\mu$m$^2$ cell from Fig. \ref{fig:fig2} (a).} (a) Current-Voltage characteristics of the readout junction: black - in the vortex-free state, red - with the trapped vortex, and blue - with an antivortex. Green dashed line represents the normal resistance, $R_n=0.61~\Omega$. (b) Demonstration of the write operation $0\rightarrow \pm 1$ by wordline pulses (without bitline currents). The readout junction resistance is shown as a function of the WL current pulse amplitude. Measurements were performed at $T=5.85$ K and $H=0$.   
    }
    \label{fig:fig3}
\end{figure*}

Table \ref{tab:tabI} summarizes measured threshold currents for all operations on this cell, performed by sending a single pulse along either WL or BL. For WL operation, we observe three threshold levels $|I_{\text{WL}}|$: $\simeq 0.15$ mA (easy), $\simeq 0.73$ mA (moderate) and $\simeq 0.87$ mA (difficult). Of those, the ``easy" corresponds to erasing, $\pm 1 \rightarrow 0$, along the track; the ``moderate" to writing $0 \rightarrow \pm 1$ along the track; and the ``difficult" to all other operations in the direction opposite to the track. Thus, the subtle track is playing a very important role. It introduces the required asymmetry, enabling a controllable realization of all four main operations along the easy path. Control in the ``difficult" regime is much worse because such currents can do any operation. 

\begin{table}[h]
    \centering
    \begin{tabular}{|c|c|c|c|c|}
    \hline
Operation & Easy & Moderate & Difficult & Hard\\
    \hline
write 1 ($0 \rightarrow 1$) & & $I_{\text{WL}} > 0$ & $I_{\text{WL}} < 0$ & BL \\
write -1 ($0 \rightarrow -1$) & &$I_{\text{WL}} < 0$ & $I_{\text{WL}} > 0$ & BL \\
erase 1 ($1 \rightarrow 0$) & $I_{\text{WL}} < 0$ & & $I_{\text{WL}} > 0$ & BL \\
erase -1 ($-1 \rightarrow 0$) & $I_{\text{WL}} > 0$ & & $I_{\text{WL}} < 0$ & BL \\
    \hline
Threshold (mA) & 0.15 & 0.73 & 0.87 & $> 1$ \\
\hline
    \end{tabular}
    \caption{{\bf Summary of single-line operations for the device from Fig. \ref{fig:fig2} (a) at $T=5.85$ K and $H=0$.} WL operations have three distinct levels. Easy ($|I_{\text{WL}}| \simeq 0.15$ mA) and moderate ($|I_{\text{WL}}| \simeq 0.73$ mA) thresholds correspond to vortex propagation along the easy track. The difficult threshold ($|I_{\text{WL}}| \simeq 0.87$ mA) - to vortex motion outside the track. For BL operations there is no easy path; therefore, all BL operations are hard ($|I_{\text{BL}}| > 1$ mA). }
    \label{tab:tabI}
\end{table}

Lorentz forces induced by BL currents are perpendicular to the track. Therefore, there is no profound asymmetry or easy path, and all BL operations are ``hard", $|I_{\text{BL}}|> 1$ mA. This is important because the passiveness of BL enables nondestructive readout. As seen from Figs. \ref{fig:fig2} (a) and (b), BL currents go through the JJs. Since fairly large BL currents do not affect the vortex state, we can nondestructively readout the state by measuring junction voltage and resistance. This is demonstrated in Figure \ref{fig:fig3} (a), which shows the current-voltage ($I$-$V$) characteristics of the readout JJ in V=0 (black), 1 (red) and -1 (blue) states. The current, sent via the BL, does not cause switching between vortex states within this bias range. 
%with large readout voltages $\sim I_{\text{BL}} R_n$ by sending large $I_{\text{BL}}$. 
Thus, the introduced asymmetry is crucial both for controllable operation and for nondestructive readout.  

\subsection{Wordline operation} 

First, we analyze the $0\rightarrow\pm 1$ write operation solely by the WL current. %for the same device. %from Fig. \ref{fig:fig2} at $H=0$. 
The experiment is done in the following manner. First, the cell is prepared in the 0-state. %by sending 
After that, a short %$\sim 20???\mu$s current 
pulse with an amplitude $I_{\text{WL}}$ is sent through the WL. The state of the device is evaluated by measuring the locking resistance, $R$, of the JJ via the BL. The probe ac current $I_{ac}\simeq 130~\mu$A is smaller than $I_c$ in the 0-state, but larger than in 1 and -1 states (see Fig. \ref{fig:fig3} (a)), so that the 0-state corresponds to $R=0~\Omega$, and 1 and -1 states to $R\simeq 0.27$ and $0.26~\Omega$, respectively.
%MORE about pulse shapes ??? The presence of four electrodes allows independent measurements of both JJs. 
Fig. \ref{fig:fig3} (b) shows the readout JJ resistance as a function of $I_{\text{WL}}$. 
%$R=0$ corresponds to the 0-state of the cell and $R\ne 0$ to the vortex (1 or -1) states. The simultaneously measured resistance of the test JJ always showed a correlated behavior (albeit with a significantly smaller response) confirming that AV was indeed located in the trap \cite{Supplem}???. 
It is seen that vortices can be written in the range $0.73~\text{mA} \lesssim |I_{\text{WL}}| \lesssim 0.87~\text{mA}$. The lower limit, $|I_{\text{WL}}| \simeq 0.73~\text{mA}$, corresponds to the ``moderate" threshold for writing along the track. As follows from Table \ref{tab:tabI}, positive $I_{\text{WL}}$ causes $0 \rightarrow 1$ switching, and negative $0 \rightarrow -1$ switching. At $|I_{\text{WL}}| > 0.87~\text{mA}$, no switching is observed. Presumably, this is caused by exceeding the ``difficult" threshold, which leads to the introduction of an opposite vortex, that annihilates with the one trapped earlier via the track.  

\begin{figure*}[t]
    \centering
    \includegraphics[width=0.9\textwidth]{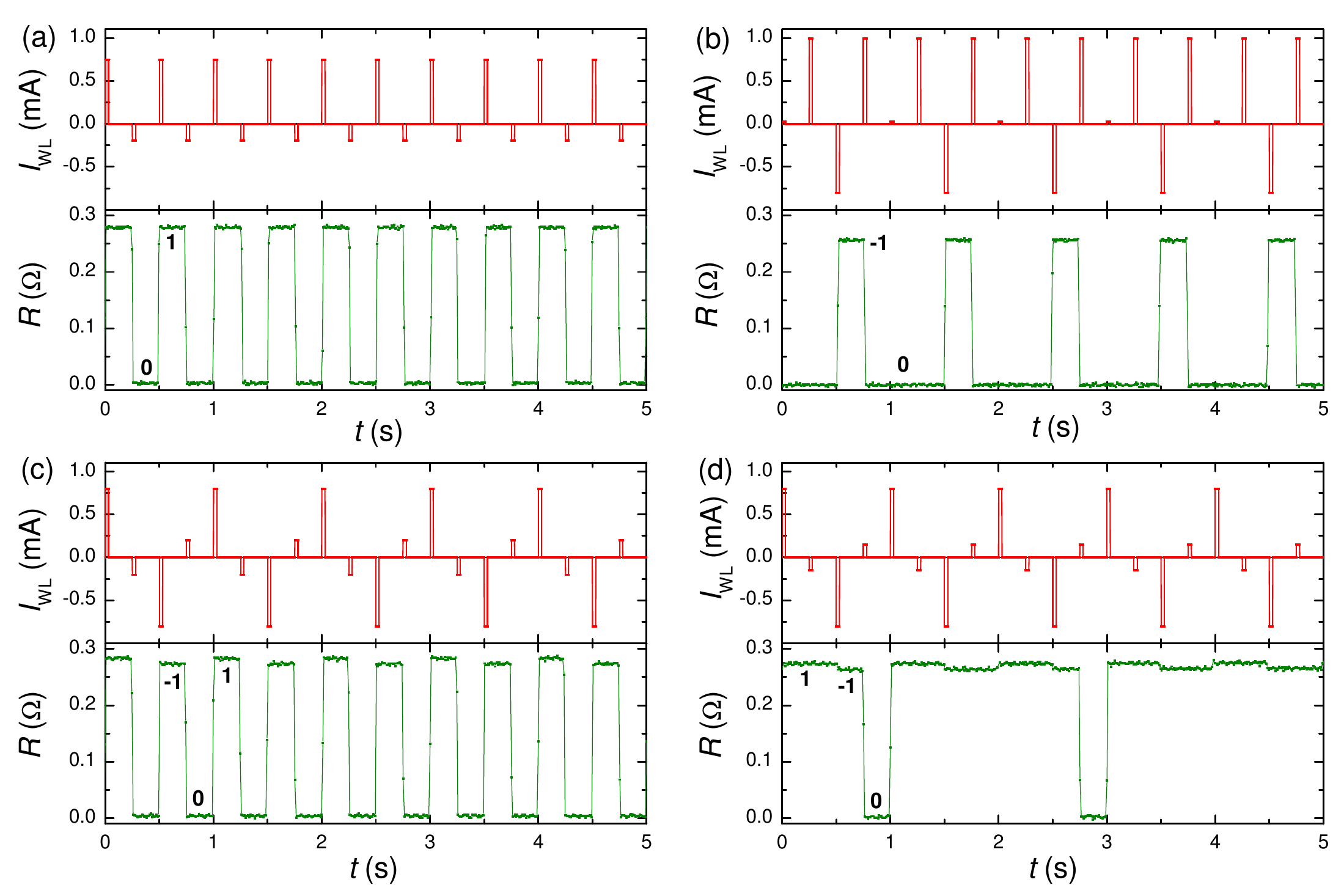}
    \caption{ {\bf Demonstration of wordline operation of the $1\times 1 ~\mu$m$^2$ cell (without bitline currents).} (a) Writing and erasing a vortex along the easy path by WL pulses 0.75 and -0.1875 mA. (b) Writing an antivortex along the easy path and erasing along the difficult path by WL pulses -0.8 and 1.0 mA. (c) Writing and erasing vortices and antivortices along the easy path by WL pulses $\pm 0.8$ and $\pm0.2$ mA. (d) The same as in (c) with WL pulses $\pm 0.8$ and $\pm0.15$ mA. In all figures, top panels show time sequences of WL pulses and bottom panels - a simultaneously measured ac-resistance of the readout junction. Measurements were performed at $T=5.85$ K and $H=0$.
    }
    \label{fig:fig4}
\end{figure*}

In Fig. \ref{fig:fig4}, we show representative examples of high-endurance operations by periodic positive/negative WL pulses. In Fig. \ref{fig:fig4} (a), a positive pulse, $I_{\text{WL}} = 0.75~\text{mA}$, slightly above the   ``moderate" threshold, writes a vortex ($0\rightarrow 1$) and a small negative pulse, $I_{\text{WL}} = 187.5~\mu\text{A}$, erases it. Both operations occur along the track. However, erasing is significantly easier than writing. %This indicates the presence of additional (not well understood) asymmetry.  

In Fig. \ref{fig:fig4} (b), a negative pulse, $I_{\text{WL}} = -0.8~\text{mA}$, %, above the moderate threshold, 
writes an antivortex ($0\rightarrow -1$). The subsequent positive pulse, $I_{\text{WL}} = 1.0~\text{mA}$, above the ``difficult" threshold, annihilates it and switches the sell into the 0-state. The next similar pulse doesn't cause switching from the 0-state, consistent with Fig. \ref{fig:fig3} (b). 

In Fig. \ref{fig:fig4} (c), a 4-pulse train with small $\pm 0.2~\text{mA}$ and moderate $\pm 0.8~\text{mA}$ pulses is applied. Here, the moderate negative/positive pulses write $-1/1$, and subsequent small positive/negative pulses erase them. All operations are achieved via the easy track. 

Fig. \ref{fig:fig4} (d) represents a similar 4-pulse train with $\pm 0.15~\text{mA}$ and $\pm 0.8~\text{mA}$ pulses. Here the small negative pulse appears to be sub-threshold and does not erase the 1-state. The subsequent moderate negative pulse causes $1\rightarrow -1$ switching, which can be considered as a sequential entrance of two antivortices, of which the first annihilates with the trapped vortex and the second stays in the trap. The small positive pulse, however, is sufficient for erasing the -1 state. The small asymmetry between 1 and -1 states, also seen from the $I$-$V$s in Fig. \ref{fig:fig3} (a), is caused either by a tiny residual field in the cryostat or by a current-induced self-field effect \cite{Golod_2022}. This allows a distinction between 1 and -1 states, as can be seen in the bottom panel of Fig. \ref{fig:fig4} (d). 

\subsection{Word and Bit-line operation}

\begin{figure*}[t]
    \centering
    \includegraphics[width=0.99\textwidth]{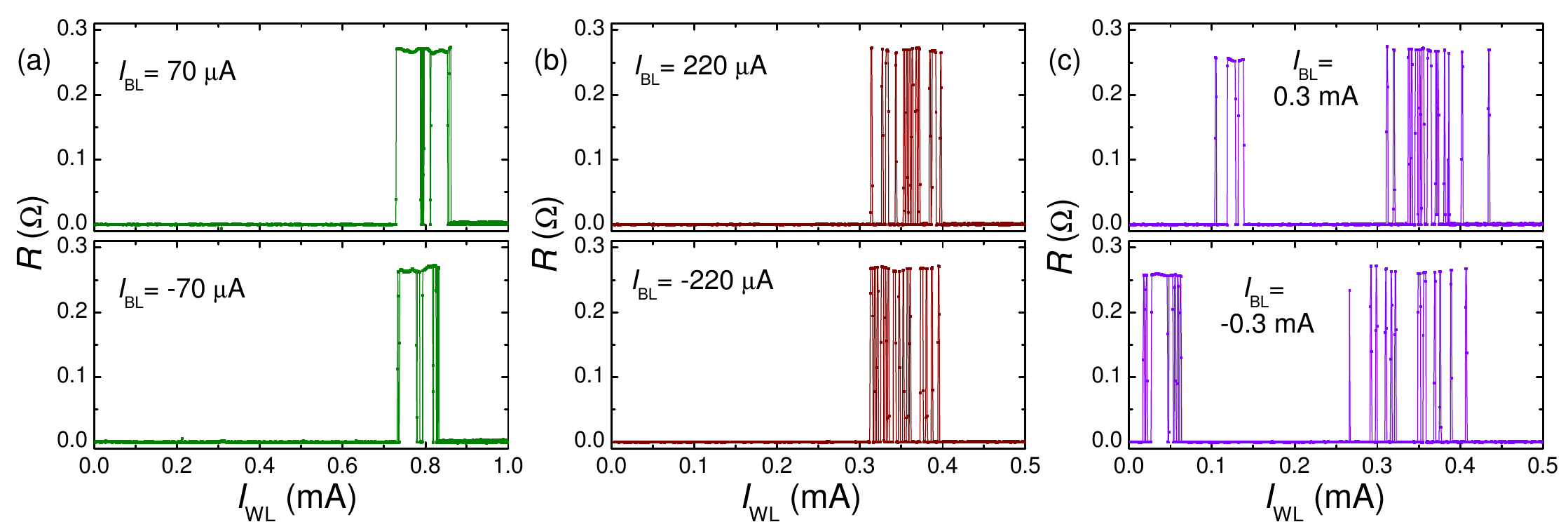}
    \caption{ {\bf Word and bit-line operation of the $1\times 1~\mu$m$^2$ cell.} The readout junction resistance is shown as a function of the positive WL pulse amplitude, $I_{\text{WL}}$, for fixed amplitudes of the BL current. (a) $I_{\text{BL}}=\pm 70~\mu$A; (b) $I_{\text{BL}}=\pm 220~\mu$A; and (c) $I_{\text{BL}}=\pm 300~\mu$A. A significant reduction of the threshold WL current is observed upon simultaneous application of the BL current. Measurements were performed at $T=5.85$ K and $H=0$.
    }
    \label{fig:fig5}
\end{figure*}

A multi-bit RAM requires selective addressing of individual cells by coincident WL and BL pulses. Figure \ref{fig:fig5} demonstrates the WL+BL write operation, $0\rightarrow \pm 1$. JJ resistance is shown as a function of the positive WL amplitude for several values of BL pulses: (a) $I_{\text{BL}}=\pm 70~\mu$A, (b) $I_{\text{BL}}=\pm 220~\mu$A, and  (c) $I_{\text{BL}}=\pm 300~\mu$A. It is seen that the threshold WL current is reducing with increasing the BL current. Remarkably, the WL+BL operation leads to a significant reduction of the total threshold current, $I_{\text{tot}}=|I_{\text{WL}}|+|I_{\text{BL}}|$: 
$I_{\text{tot}} \simeq 730~\mu$A for $I_{\text{BL}}= 0$ (see Fig. \ref{fig:fig3}) and $\pm70~\mu$A; $I_{\text{tot}}\simeq 520 ~\mu$A for $I_{\text{BL}}= \pm 220~\mu$A; and $I_{\text{tot}}\simeq 320-400 ~\mu$A for $I_{\text{BL}}= \pm 300~\mu$A. The latter is approximately two times smaller than for a solo WL operation. Thus, the cooperative effect of WL and BL pulses is significantly better than a sum of the two currents. This is very good because it can greatly improve the selectivity of addressing individual cells in a multi-bit AVRAM.    

\begin{figure*}[t]
    \centering
    \includegraphics[width=0.9\textwidth]{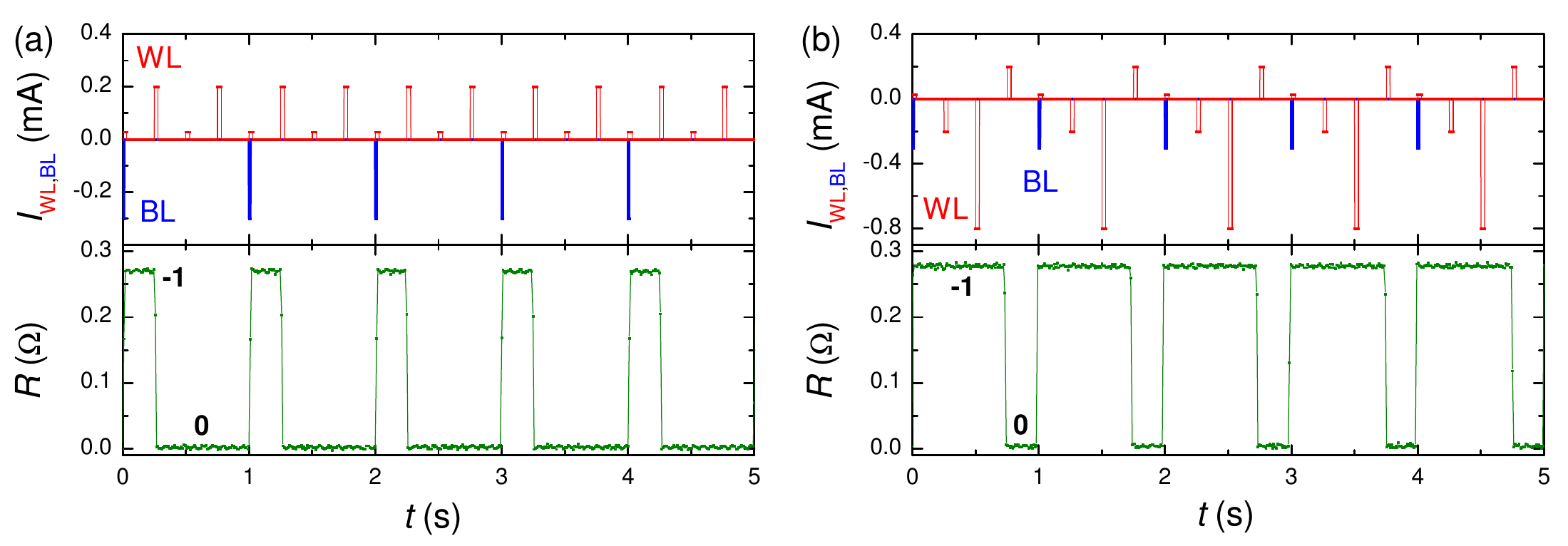}
    \caption{ {\bf Demonstration of a simultaneous word and bit-line operation of the $1\times 1~\mu$m$^2$ cell.} (a) Writing an antivortex by combined pulses $I_{\text{BL}}= -300~\mu$A and $I_{\text{WL}}= 29~\mu$A; and erasing by a WL pulse $200~\mu$A. (b) Similar write and erase pulses as in (a) with additional intermediate WL pulses. It is seen that the negative WL pulses do not change the -1 state because they tend to write an antivortex and the trap is already occupied. In all figures, top panels show time sequences of WL (red) and BL (blue) pulses. Bottom panels show a simultaneously measured ac-resistance of the readout junction. Measurements were performed at $T=5.85$ K and $H=0$.
    }
    \label{fig:fig7}
\end{figure*}

Fig. \ref{fig:fig7} demonstrates switching dynamics with WL and BL pulses. Panel (a) shows low-threshold switching. Here a small $I_{\text{BL}}= -300~\mu$A together with a very small $I_{\text{WL}}= 29~\mu$A causes $0\rightarrow -1$ switching and a small $I_{\text{WL}}= 200~\mu$A erases it, $-1\rightarrow 0$. Subsequent WL-only pulses with the same amplitudes, 29 and 200 $\mu$A, do not affect the 0-state. In panel (b), similar write and erase pulses are applied. Negative WL-only pulses -200 and -800 $\mu$A, applied in the -1 state, do not affect the state. Even larger BL-only pulses, $|I_{\text{BL}}|>1$ mA, can be applied without changing the state of the device (see Table \ref{tab:tabI}). This illustrates that the smallness of the combined WL+BL current does not affect the robustness of nondestructive readout.  

\section{Discussion}

%\subsection{Scalability}
We demonstrated that the planar AVRAM cell can be straightforwardly scaled down to $\sim 1\times 1~\mu$m$^2$. Furthermore, threshold currents for miniaturized cells are significantly reduced, compared to larger cells, studied in Ref. \cite{Golod_2015}. We observe that the threshold WL current is approximately proportional to the width of the WL electrode. This is expected because vortex motion occurs when the transport current density exceeds the depinning current density. Consequently, upon miniaturization, the threshold current scales down proportional to the size.
Therefore, AVRAM cells do not suffer from the problem of threshold current enhancement upon miniaturization, as SQUIDs in RSFQ. This proves that AVRAM cells are indeed scalable. %to the size, determined by the London penetration depth, $\lambda_L \sim 100$ nm. 

The area of the presented $\sim 1\times 1~\mu$m$^2$ cell is already $\sim 100$ times smaller than for the state of the art RSFQ cells \cite{Semenov_2019,Chen_2020}. But this is not the ultimate limit. 
Miniaturization of the ARVAM cell is limited by the stability of AV in a mesoscopic superconducting electrode. In order to work as a memory bit, the vortex should be persistent %(metastable) 
in the trap at zero magnetic field. Metastable vortex configurations in mesoscopic superconductors have been intensively studied and are well understood \cite{Bezryadin_1995,Geim_1997,Berdiyrov_2003,Chibotaru_2005,Milosevic_2009,Vodolazov_2011,Oliveira_2014}. 
The trap plays an important role in stabilizing the vortex. It changes the electrode topology and creates a sharp pinning potential, which stabilizes otherwise energetically unfavorable $\pm 1$ vortex states at zero field. The tendency of the vortex to spontaneously jump out of the electrode can be viewed as the consequence of an attractive interaction with an image antivortex outside the electrode \cite{Golod_2019}. From Fig. \ref{fig:fig2} (a) it can be seen that for the studied $1\times 1~\mu$m$^2$ device the nearest to the trap edge corresponds to the readout JJ, 240 nm away. Apparently, this distance is sufficient for preventing a spontaneous vortex exit. Consequently, the cell can be straightforwardly minituarized to twice this size, i.e., to $\sim 500\times 500$ nm$^2$. 

Generally, vortices can exist in superconductors with sizes down to the coherence length, $\xi$. However, the smaller is the size, the larger is the relative energy cost of a vortex at zero field. %magnetic field, required for stabilizing the vortex.
Our sputtered Nb films have a short coherence length of $\xi (0) = 14$ nm and a London penetration depth, $\lambda_L(0)\simeq 100$ nm \cite{Zeinali_2016}. Therefore, we anticipate that further miniaturization down to $\sim 100$ nm should be feasible.

%\subsection{Conscious geometrical design}

We have emphasized that a specific geometrical asymmetry is needed for controllable vortex manipulation. This is particularly important for submicron AVRAM cells, which are in the mesoscopic limit. %, determined by $\lambda_L$. 
Geometry (sizes and shapes) is playing a crucial role for vortex states in mesoscopic superconductors \cite{Bezryadin_1995,Geim_1997,Berdiyrov_2003,Chibotaru_2005,Milosevic_2009,Vodolazov_2011,Oliveira_2014}. Proper design of the cell geometry (the trap and the electrode) would be needed for optimization of AVRAM. 

In this work we introduced several small but important improvements in the cell design, compared to the initial prototype \cite{Golod_2015}. First, the film structure was reduced to a single Nb film (compared to a bilayer superconductor/normal metal or superconductor/ferromagnet) and the readout junction structure became of a variable-thickness-bridge type (compared to proximity-coupled SNS or SFS JJs). Apart from a bare simplification, this leads to a major enhancement of the $I_c R_n$ product, reaching almost 1 mV at low $T$ \cite{Grebenchuk_2022}. This results in a proportional enhancement of the readout voltage \cite{Golod_2022}, which is very important for device applications. 

Most importantly, we have consciously engineered geometrical asymmetry. An easy track for vortex manipulation was made at one side of the trap. This led to improved controllability and reduced threshold currents. The analysis in Table \ref{tab:tabI} confirms that the track works and vortices indeed follow this route.
The achieved threshold currents, as low as 150 $\mu$A, are optimal for practical AVRAM. Further reduction would jeopardize the nondestructive readout and lead to smaller readout signals. 
Other geometrical features that help to guide vortex motion were also suggested in the literature \cite{Vodolazov_2011,Oliveira_2014} and could be implemented, if necessary. 
 
%\subsection{Perspectives of multibit AVRAM}

%The main focus of our work was made on detailed analysis of the WL and BL operation, see Figs. 3-6. In Fig. \ref{fig:fig7} we demonstrated high-endurance WL+BL operation with sub-mA currents. Most importantly, we observed that the cooperative WL+BL effect significantly exceeds the bare sum of WL and BL currents. As demonstrated in Fig. \ref{fig:fig5}, the combined WL+BL threshold current can be reduced by a factor two, while the WL current in the bottom panel of Fig. \ref{fig:fig5} (c) is reduced by more than an order of magnitude, compared to the solo WL current, presented in Fig. \ref{fig:fig3} (b). This is very good for the selectivity of addressing individual cells in a multibit RAM. The smaller required WL and BL currents for the combined WL+BL operation reduce chances of affecting other cells at the same WL. Nevertheless, large crosstalking will be unavoidable in an interconnected mesh of cells. Therefore, a multibit AVRAM would require additional, electrically isolated control lines for each bit line. 

%Furthermore, the write current should be kept small ($\ll 1$ mA) in order to prevent overheating and the footprint of the cell should be $<1~\mu$m$^2$ in order  

%$S_V=5$ nV$/\sqrt{\text{Hz}}$, $V=50~\mu$V, $R_n=0.5~\Omega$, $\tau = (S_V/V)^2\simeq 10$ ns, $E=V^2\tau/R_n =S_V^2/R_n \simeq 50$ aJ (atto-Joule) 

%Current-locked arrays \cite{Grebenchuk_2022}

\section{Conclusions}
We have studied single-bit AVRAM cells and arrived to three main conclusions.  

(i) The planar AVRAM cell design is straightforwardly scalable to submicron sizes. The threshold current is reducing upon miniaturization, approximately proportionally to the size, thus obviating the problem of SQUID-based RSFQ cells. Robust, non-volatile operation at zero magnetic field was demonstrated for a $\sim 1\times 1\mu$m$^2$ cell. Its area is approximately 100 times smaller than for the state of the art RSFQ cells. 

(ii) We emphasized the importance of a conscious geometrical design of the cell geometry. A specific geometrical asymmetry (an easy track) was introduced in our cells. It enabled controllable and comprehensible operation of the cell.

(iii) Word and bit line operation exhibits a profound cooperative effect, reducing the total threshold current. 
The smaller combined WL+BL currents reduce chances of affecting other cells at the same WL.
This can improve the selectivity of cell addressing in a multibit RAM. 

We conclude that planar AVRAM cells are promising candidates for the creation of VLSI-compatible superconducting memory. The achieved one square micron area is an important milestone and a significant step forward in this direction.

\end{document}